 \documentclass[preprint2]{aastex}

\shorttitle{Djorgovski et al.}
\shortauthors{Collapsed Cores in Globular Clusters}

\begin{document}


\title{Infrared Light Curves of Mira Variable Stars\\ from
COBE DIRBE Data}


\author{Beverly J. Smith}
\affil{Department of Physics and Astronomy, East Tennessee State University,
    Box 70652, Johnson City, TN  37614}
\email{smithbj@etsu.edu}

\author{David Leisawitz}
\affil{Goddard Space Flight Center}
\email{leisawitz@stars.gsfc.nasa.gov}

\author{Michael W. Castelaz}
\affil{Pisgah Astronomical Research Institute, 1 PARI Drive, Rosman, NC  28772}
\email{mcastelaz@pari.edu}

\author{Donald Luttermoser}
\affil{Department of Physics and Astronomy, East Tennessee State University,
    Box 70652, Johnson City, TN  37614}
\email{lutter@etsu.edu}



\begin{abstract}
We have used the COBE DIRBE database
to derive near- and 
mid-infrared
light curves for a well-defined sample of 38 infrared-bright 
Mira variable stars, and compared with optical data from the AAVSO.
In general, the 3.5 $\mu$m and 4.9 $\mu$m DIRBE bandpasses
provide the best S/N light curves, with S/N decreasing
with wavelength at longer wavelengths.  At 25 $\mu$m, 
good light curves are only available for $\sim$10$\%$ of our stars, and
at wavelengths $\ge$ 60 $\mu$m, extracting high quality
light curves is not possible.
The amplitude of variability is typically less 
in the near-infrared than in the optical, and less in
the mid-infrared than in the near-infrared, with decreasing
amplitude with increasing wavelength.  On average, there are
0.20 $\pm$ 0.01
magnitudes variation at 1.25 $\mu$m and 
0.14 $\pm$ 0.01 magnitudes variation at 4.9~$\mu$m
for each magnitude
variation in V.
The observed amplitudes are consistent with results
of recent theoretical models of circumstellar
dust shells around Mira variables.
For a few stars in our sample, we find 
clear evidence of 
time lags between the optical and
near-infrared maxima of 
phase $\sim$ 0.05 $-$ 0.13, with no lags in the minima.
For three stars, mid-infrared maximum appears to occur slightly before
that in the near-infrared, but after optical maximum.
We find three examples of secondary
maxima in the rising portions of the DIRBE light curves,
all of which have 
optical counterparts in the AAVSO data, supporting the hypothesis
that they
are due to shocks rather than newly-formed dust layers.
We find no conclusive evidence for rapid
(hours to days) variations in the infrared brightnesses of these stars.

\end{abstract}




\section{Introduction}

Current theories of stellar evolution indicate that 
low mass ($\le$8 M$_{\sun}$) stars pass through the 
pulsating Mira star phase at the end of their time as Asymptotic
Giant Branch stars \citep{w97, o99}.   
Although the time that stars spend as 
Mira variables is relatively short 
($\sim$6 $\times$ 10$^4$ years;
\citet{w90}), this stage is very important because these stars have 
very high rates of
mass loss (up to 10$^{-4}$ M$_{\sun}$ 
yr$^{-1}$; \citet{km85}).
This means that they return large quantities of
carbon, oxygen, and nitrogen-enriched gas to the 
interstellar medium \citep{s94}.
At present, the mass loss process, the properties of the circumstellar
envelopes of these stars, and how
these relate to the variability of the star 
are still not very well-understood.

To better understand these processes, detailed studies 
of Mira variables
at
infrared wavelengths are helpful.  In the
near-infrared, photospheric emission dominates,
while at wavelengths longer than $\sim$3~$\mu$m 
emission from dust in the circumstellar envelope becomes
important \citep{wn69}.
Infrared observations have been used to constrain models of the
circumstellar envelopes of Mira variables and to estimate mass
loss rates from these
stars \citep{j86, b87, o89a, o89b, apv93, lw98}.
These studies, however,
are based on time-averaged 
broadband fluxes from the Infrared Astronomical
Satellite (IRAS) \citep{j86, b87,
apv93, lw98}
or time-averaged 8 $-$ 23 $\mu$m IRAS 
Low Resolution Spectrometer (LRS) spectra
\citep{o89a, o89b},
and have not investigated mid- or far-infrared
variability of these stars.

To better understand mass loss in these stars, infrared monitoring
is helpful.  Such studies have
revealed
variations in the infrared spectral energy distributions of
Miras with time.
Broadband infrared observations indicate
that the amplitude of variation
tends to decrease with wavelength
from the optical through the mid-infrared \citep{lw71, h74, c79,
bcm96, lb92, lb93, l96}.
Spectroscopic observations with IRAS
\citep{ls92, ha94, l96},
ground-based telescopes \citep{c97, mgd98}, and the Infrared
Space Observatory (ISO)
\citep{o97} showed that at least for some Miras 
the silicate features tend to
decrease in flux from light curve maximum to minimum.
There is also evidence from IRAS data for 
far-infrared variations of Miras 
\citep{ls92, l96, som93},
however, the IRAS light curves are poorly-sampled, so
variations at these wavelengths are not well-studied.

Infrared monitoring of Mira variables
has also revealed wavelength-dependent phase lags.
The maxima
at 1~$\mu$m 
\citep{pn33, lw71, b73} and 2.7 $\mu$m \citep{m77}
typically lag those at V by about 0.1 phase, while
no lags are seen in the minima.
At present, phase lags at longer wavelengths are
not well-determined.
There have also been a few reported observations of
large-amplitude (0.5 $-$ 1 magnitudes)
very short-term (hours to days) variations in the visual
\citep{mt95, dl98} and near-infrared \citep{sw79b, gh01}
light curves of Mira variables, but such variations have
not yet been reported at longer wavelengths.

The previously-cited
studies generally provide long-wavelength infrared light
curves for only a relatively small sample of stars, and
generally sample only a fraction of the
light curve of these stars.
To obtain more complete infrared light curves for a larger sample
of Mira stars,
we have used data from the 
Diffuse Infrared Background Experiment (DIRBE) 
instrument \citep{h98}
on 
the Cosmic Background Explorer (COBE) spacecraft \citep{b92}.
To date, little research has been done on infrared
variability of stars using DIRBE data, in spite of its suitability
for such studies.

The 
COBE satellite was launched in November 1989, and
the DIRBE instrument functioned at cryogenic temperatures 
from 11 December 1989 until 21 September 1990.  
This 10 month timescale is suitable for searching for variations
in long period variables.
In the course of a week, a typical point on the sky was observed 10-15 times
by DIRBE,
and over the course of the cryogenic mission, it was 
observed about 200 times \citep{h98}.
The temporal coverage of DIRBE varies with sky position.  Some parts
of the sky had relatively complete coverage; for other regions,
there were gaps in the coverage of several months \citep{h98}.
DIRBE operated at 10 infrared wavelengths
(1.25, 2.2, 3.5, 4.9, 12, 25, 60, 100, 140, and 240 $\mu$m),
providing a nearly complete infrared spectrum.  This allows
us to tie stellar phenomena, as seen in the visual and near-infrared,
to circumstellar emission, seen at longer wavelengths.
The beam size of the DIRBE instrument, 0.7 degrees, although
large, is still sufficiently small that individual infrared-bright
stars can be discerned.

We have used DIRBE data to extract 1.25 $\mu$m $-$ 240 $\mu$m
light curves for 38 Mira variables and compared them to 
available optical data.  We have investigated
the shapes of the light curves, amplitude variations, and 
phase lags, and have compared these results to recent
models of circumstellar dust emission from Miras.

\section{Sample Selection and Infrared Light Curve Derivation}

Our stars were selected from the 12 $\mu$m flux-limited sample
of variable stars 
described by \citet{sp98}
and \citet{slp98}.
This sample was 
obtained by cross-referencing the General Catalogue of
Variable Stars \citep{k85}
with the \citet{PSC}.
We selected a subset of this sample, with an IRAS 12 $\mu$m 
flux density cutoff of
235 Jy 
and classification as Mira stars.  
A total of 38 stars are included in
our set (Tables 1 and 2).  Eight of these are classified
as carbon-rich Mira variables (Table 1), while 30 are oxygen-rich (Table 2).
In Tables 1 and 2, we provide the positions,
the period of variability, the IRAS 12 $\mu$m
magnitude, the optical spectral type, and the dust spectral type (based
on the IRAS LRS spectra).
The positions are from the SIMBAD database; the rest of 
the information is from \citet{sp98} or \citet{slp98}.

Infrared light curves were derived for these
stars from the DIRBE data using two methods.  We first
extracted light curves
from 
the COBE DIRBE Weekly Sky Maps.  These are weekly-averaged intensity
maps of the sky in the ten DIRBE wavelength bands.
The stellar photometry was determined by multiplying the surface
brightess of the Weekly Sky
Maps at the stellar position by
the DIRBE beam solid angle, after sky subtraction.
For sky values, we used the
average of the fluxes at four positions 2$^{\circ}$ N, S, E, and W in
equatorial coordinates.
Flux
uncertainties were obtained by using
the quoted DIRBE photometric uncertainties
combined in quadrature with the dispersion in the four
sky measurements.
At 1.25 $\mu$m $-$ 240 $\mu$m,
respectively,
the average noise levels in these light curves are 
35 Jy, 40 Jy, 35 Jy, 20 Jy, 120 Jy, 155 Jy,
140 Jy, 460 Jy, 2850 Jy, and 1250 Jy.

The second method of obtaining light curves
used the DIRBE Calibrated Individual Observations (CIO) data files.
The CIO database contains the calibrated individual 1/8th second
samples taken in science-survey mode during each day of the
cryogenic mission.
For all scans that passed within 0.3$^{\circ}$ of the target 
position, a linear baseline was
fit to the sections $\pm$1.35$^{\circ}$ $-$ 2.25$^{\circ}$ from the point
of closest approach.  The point source photometry was obtained
by subtraction of this baseline and correcting for DIRBE beam response. 
The uncertainties in the point source photometry are calculated as
the quadrature sum of the rms noise of the baseline,
an error due to positional uncertainties
of 1$'$ in both the in-scan and cross-scan directions,
an error due to short-term detector gain variations, and 
signal-dependent detector noise.
For the light curves derived
from the CIO data, average noise levels per datapoint are 
25 Jy, 20 Jy, 20 Jy, 10 Jy, 30 Jy, 55 Jy, 320 Jy, 765 Jy,
4800 Jy, and 2750 Jy for the DIRBE 1.25 $-$ 240 $\mu$m filters, respectively.

Datapoints with large uncertainties
(greater than three times the average uncertainty) were 
eliminated
from the dataset, since
anomalously large errorbars were likely caused by cosmic ray hits or 
the presence of 
a nearby companion in the DIRBE scan.
Careful comparison of 
individual DIRBE scan locations with the IRAS Point Source Catalog (1988)
and the all-sky near-infrared catalog developed by Smith (1995) for position
reconstruction for the Infrared Telescope in Space (IRTS) mission
\citep{m96, wheat97} shows that
when an infrared-bright companion was present in the baseline portion
of the scan, 
the derived stellar photometry was sometimes 
discrepant with large uncertainties.

These two methods provided consistent 
light curves, with flux densities that agree within 
15$\%$, 4$\%$, 12$\%$, 2$\%$, 6$\%$, 3$\%$, and 12$\%$,
for the 1.25 $\mu$m - 60 $\mu$m bands, respectively, when averaged
over our light curves.
As noted in the DIRBE Explanatory Supplement \citep{h98}, point source
photometry from the Weekly Sky Maps may suffer a small systematic
offset when compared to the CIO data.
The CIO data are believed to be
more accurate for point source photometry \citep{h98},
so for the following
analysis
we focus on the CIO light curves.
Note that for the six shortest wavelengths, the
uncertainties in the CIO light curves are 
less than those of the photometry derived from the Weekly
Sky Maps, in spite of the shorter time coverage per data point.
This provides a second reason for using the CIO
light curves rather than the light curves from the Weekly Sky Maps.

Note that the uncertainties used in this paper do not include absolute flux
calibration errors, and  
that color corrections were not made to the
DIRBE
flux densities.
The color corrections are typically quite small, particularly
at the shorter DIRBE wavelengths.
For the stars in our sample,
the color correction tables in \citet{h98} give
estimated color correction factors (F$_{observed}$/F$_{true}$)
that
range from
0.99 $-$ 1.03 at 1.25 $\mu$m, 0.99 $-$ 1.03 
at 2.2 $\mu$m, 0.98 $-$ 1.01 at 3.5 $\mu$m,
0.99 $-$ 1.01 at 4.9 $\mu$m, 0.89 $-$ 1.1 at 12 $\mu$m, 
and 0.5 $-$ 2.8 at 25 $\mu$m.
The range quoted for each wavelength reflects the
uncertainty associated with the shape of the emissivity law
($\epsilon \propto \nu^{1 - 2}$), the fact that the spectral
energy distribution of each star
is variable in time, and the fact that each star has a slightly
different spectral energy distribution.




\section{Optical Data}

For the period
of the COBE mission, 23 out of our 38 stars have well-sampled visual (V)
light curves 
available from the American Association
of Variable Star Observers (AAVSO) dataset \citep{m2000}.
These stars are identified in the last column of Tables 1 and 2.
These optical
light curves are plotted along with the infrared data in Figures 1 and 2.

\section{Results}

\subsection{Overview of the Light Curves}

In Figure 1 we present
plots of flux density vs.\ time 
for the six shortest wavelength DIRBE bands
for six stars with relatively
complete DIRBE light curves.
In Figure 1, we also include the visual light curves when available, as well
as
the 2.2~$\mu$m to 12~$\mu$m flux ratio as a function
of time.  For RW Vel, which does not have optical data available,
we also plot F$_{4.9~{\mu}m}$/F$_{2.2~{\mu}m}$ vs.\ time.
In Figure 2, we provide 4.9 $\mu$m light curves for the remaining
31 stars in our sample, along with optical light curves when available.
These same data are displayed in a different manner in Figure 3, where
we plot the optical-infrared spectral energy distributions for
all 38 stars at one or more
times during the COBE mission vs.\ wavelength.

In general, the 3.5 $\mu$m and 4.9 $\mu$m light curves have
the best S/N.  
At 3.5 $\mu$m and 4.9 $\mu$m, respectively, on average the stars 
have 18.0 and 17.1 5$\sigma$ weekly-averaged flux measurements from the 
Weekly Sky Maps (out of a total possible of 41 weeks).  
In the CIO light curves, at 3.5 $\mu$m, the average number of datapoints 
is 490, and at 4.9 $\mu$m, 500 points 
are available on average.  Almost complete 4.9 $\mu$m light curves 
($\ge$30 weeks of 5$\sigma$ data) are available for five stars in 
the sample, while 31 stars (82$\%$ of the sample) have at 
least 10 5$\sigma$ datapoints available.  Beyond 4.9 $\mu$m, 
the S/N drops with increasing wavelength.
At 12 $\mu$m, for example,
out of the 38 sample stars only 14 (37$\%$) have 
more than ten datapoints detected
at the 5$\sigma$ level or better in the Weekly Maps.
By 25 $\mu$m, this number is down to nine stars (24$\%$), 
and at the four longest DIRBE
wavelengths, no stars have more than ten 5$\sigma$ 
weekly datapoints in their Weekly Map
light curves.

The stars in our sample are expected to be among the
brightest in the infrared, so these
statistics show the limitations of DIRBE data in studying
infrared variability of unresolved
sources.  At 60 $\mu$m and longer wavelengths, the 
signal-to-noise ratio
is too poor to obtain light curves using DIRBE data.
At 25 $\mu$m, good light curves (20 weeks at 5$\sigma$) are available
for only four out of our 38 stars (11$\%$).

\subsection{Infrared Variability of our Sample Stars from
DIRBE}

In many of the light curves
shown
in Figures 1 and 2, variations are clearly apparent in the DIRBE data.
Variations are also clearly evident in the spectral energy distributions
plotted in Figure 3.
The best wavelengths at which to observe variability are 3.5 $\mu$m
and 4.9 $\mu$m, where 
30 and 31 stars, respectively, show
$\ge$3$\sigma$ variability
in the Weekly Maps. At 25 $\mu$m, only 3 stars vary more 
than 3$\sigma$ in the weekly data.

As an approximation
based on the general appearance of the
light curves in Figure 1,
when possible
we fit the infrared light curves (in magnitude
units)
to sine functions
to estimate amplitudes and dates of maxima (Table 3).
Such fits were only possible for three stars in our
sample: R Hor (Figure 1a), RW Vel (Figure 1b), and R Cas (Figure 1c).
The rest of the light curves were too incomplete
or too peculiar to be accurately fit to a sine function.
Even for R Hor, RW Vel, and R Cas, these fits should be
taken as approximations, since the light curves are not
perfect sine functions.  These fits, however, provide useful 
estimates of the amplitudes of variation and the JD
dates of maxima as a function of wavelength (Table 3).
Since we have less than a full pulsation cycle of infrared
data for these
stars, we were unable to obtain a good fit 
for the period of variability
from the DIRBE data alone.
We therefore fixed the period
to be equal to that in the optical, and fit the DIRBE
data for the amplitude
and time of maximum.
For R Hor and R Cas, we obtained the optical period by fitting
the AAVSO data for JD = 2447500 $-$ 2448700 ($\sim$3 pulsation cycles)
to sine functions.
We found periods of 416.5 days and 439.7 days for R Hor and R Cas,
respectively, close to the values given in \citet{sp98} (Table 2).
For RW Vel, which does not have AAVSO data available for this
time period, we simply used the period given in \citet{sp98} of
443.1 days (Table 2).

The stars in our sample with almost complete DIRBE light curves (Figure 1)
can be divided into two groups: those with approximately
sinusoidal infrared light curves (R Hor, R Cas, and RW Vel),
and those which show a plateau, or inflection point, in the rising
portion of their light curves (R Car, T Cep, and S Cep).
Note that all three of the stars with sinusoidal light curves
are oxygen-rich stars, while two stars with 
inflection points are oxygen-rich (R Car and T Cep)
and one (S Cep) is a carbon star.

The best example of a secondary maximum is seen in
the light curves of R Car (Figure 1b) at JD $\sim$ 2448045 (phase $\sim$ 0.75), 
where it
is visible 
at all wavelengths between 1.25~$\mu$m and 12~$\mu$m,
and maybe at 25 $\mu$m as well, with a duration of $\sim$ 40 days 
(Figure 1).
These secondary maxima show an increase in brightness of $\sim$15 $-$ 
40$\%$ relative to that expected from a sinusoidal light curve.
This inflection point is also visible in the optical light curve,
at a similar level ($\sim$0.4 magnitudes). 

The lower S/N light curves of T Cep (Figure 1c) show weak inflection
points at JD $\sim$ 2448020 
at 1.25 $\mu$m, 2.2 $\mu$m, 3.5 $\mu$m,
and 4.9 $\mu$m, with a duration of $\sim$ 90 days.
S~Cep (Figure 1a) has possible inflection points 
between JD $\sim$ 2447980$-$2448020 in the four shortest wavelength 
DIRBE bands.
In both cases, optical counterparts are indicated by
the AAVSO data, with $\Delta$mag(V) $\sim$ 0.4.

We also searched the CIO light curves for large amplitude,
short-term (2 hours to 10 days) variations
in brightness, but found no conclusive detections greater than
0.2 magnitudes at the 3$\sigma$ level.
A handful of datapoints with very discrepant fluxes were noted in the CIO
light curves, however, these were always individual isolated points,
with discrepant fluxes at only
one wavelength, thus their anomalous fluxes
are likely due to cosmic ray hits
rather than stellar variations.
We note that in the CIO data, the typical time between
successive observations of the same star is 
$\sim$9 hours, so these data are not particularly
sensitive to extremely short variations.

\subsection{Wavelength-Dependent Phase Lags}

We used our light curves to search for 
phase lags of the maxima as a function of wavelength.
Five of our stars (R Hor, R Cas, R Cen, R Aql, and R Aqr), 
all of which are oxygen-rich, 
have both optical and infrared data available near maximum light.
All five stars show hints 
of an infrared delay relative to V.

The best example is R Hor (Figure 1a), which
clearly shows a delay of the infrared relative to V
at 1.25 $\mu$m, 2.2 $\mu$m, 3.5 $\mu$m, and 4.9 $\mu$m,
and perhaps at 12 $\mu$m and 25 $\mu$m as well.
The optical light curve peaks at JD $\sim$ 2447930,
while the infrared light curves are still rising until at least 
JD $\sim$ 2447960,
after which there are gaps in the DIRBE data
until JD $\sim$ 2447990.  These gaps
are followed by a decrease in the infrared brightness.
This phase lag can also be seen in Figure 3d, the spectral energy
distribution plots.  The optical flux was brighter
in the JD 2447897 spectrum than in the JD 2447950 spectrum, 
while the infrared fluxes were
brighter at JD 2447950.

The existence of this lag is supported by our sinusoidal
fits to the light curves (Table 3), which show that
the 1.25 $\mu$m maximum
followed that at V by $\sim$53 days (0.13 phase) and the
2.2 $\mu$m maximum lagged V by $\sim$65 days (0.16 phase).
In Figure 4a, for the three stars for which we obtained sinusoidal fits to
the light curves, we plot the phase lag relative to V
as a function of wavelength.
For RW Vel, for which we have no optical data, we assumed a 1.25 $\mu$m
lag relative to V of 0.1 phase.

R Cas also 
shows evidence for an infrared to V time lag (Figure 1c; Figure 4).
In the visible, R Cas was brightest at
JD $\sim$ 2447930, while the
1.25 $\mu$m $-$ 25 $\mu$m emission was still increasing 10 days later,
at JD $\sim$ 2447940 (phase $\sim$ 0.02), 
after which there is a gap in the DIRBE
coverage.  
The best-fit sine curves (Table 3) show a 1.25 $\mu$m lag relative
to V of 38 days (0.09 phase) and a 2.2 $\mu$m lag relative to V
of 47 days (0.11 phase).

Our data suggest not just offsets between optical to infrared maxima, but
also offsets between the DIRBE
wavelengths.  For two of our stars with relatively
complete, high S/N DIRBE light curves, R Hor and R Cas,
the shapes of the observed light
curves differ from wavelength to wavelength.
In particular, comparison of the 2.2 $\mu$m, 4.9 $\mu$m,
and 12 $\mu$m light
curves hint that the 
4.9 $\mu$m and 12 $\mu$m maxima for these stars may have occurred before
the 2.2 $\mu$m peaks (but after the optical maxima).  
The existence of these offsets is supported by our sinusoidal
fits to the data (Table 3; Figure 4).
For R Hor, the best fit maxima at 1.25 $\mu$m to 3.5 $\mu$m are 
at JD = 2447993 $\pm$ 6 days (0.13 $-$ 0.16 phase), while
those at 4.9 $\mu$m to 25 $\mu$m were 30 $-$ 50 days earlier (0.06 $-$ 0.12
phase).
For R Cas, the maxima at 1.25 $\mu$m to 3.5 $\mu$m occurred at JD =
2447999 $\pm$ 6 ($\sim$0.10 phase), 
while the 4.9 $\mu$m to 25 $\mu$m peak is at JD =
2447979 $\pm$ 3, 20 days earlier ($\sim$0.05 phase).

Of the stars with no optical data available, the most complete
infrared light curves are available for RW Vel (Figure 1b).
For this star, the maxima at 1.25~$\mu$m, 2.2~$\mu$m, and 3.5~$\mu$m
occurred at JD = 2448023 $\pm$ 9, while the 4.9 $\mu$m maximum
happened 
40 days earlier (0.09 phase) (Table 3).
This is also visible in the 
F$_{4.9~{\mu}m}$/F$_{2.2~{\mu}m}$ time plot (Figure 1b).
Other stars with possible phase lags include 
R Cen,
R Aqr, and R Aql, whose 1.25 $\mu$m maxima appear to lag those at V by
$\sim$30 days (0.05 $-$ 0.1 phase).
These are uncertain, however, because of incompleteness in
the light curves.

In addition to a mid-infrared/near-infrared phase shift, there may
be smaller offsets between the three near-infrared bands.
In particular, note that for the three stars with
sinusoidal fits to their light curves (Table 3; Figure 4), 
the 2.2 $\mu$m lags are always 
larger than those at 1.25 $\mu$m and 3.5 $\mu$m.  This result,
however, is uncertain because these light curves are not perfect
sine curves.

For six stars in our sample, 
there are both optical and infrared
data available for a minimum in the light curve: R Cen, V Cyg,
T Cep, R Lep, R Aqr, and R Car.
In all of these cases, there is no evidence for a wavelength-dependent
time delay.
Of the six, the highest S/N data are those of R Car (Figure 1b), 
which provides
a 3$\sigma$
upper limit to the time lag relative to V for 1.25 $\mu$m $-$ 25 $\mu$m
of $\sim$50 days (phase $\sim$ 0.16).
The data for R Lep (Figure 2a)
provide a 3$\sigma$ upper limit to a possible 4.9 $\mu$m to V
time lag of 20 days (phase $\sim$ 0.05), 
however, strong limits are not available for
the other DIRBE wavelengths.

\subsection{Amplitude Variations with Wavelength}

Except for the phase lags noted above, in general
the DIRBE light curves trace the optical light curves, but
with smaller amplitudes of variation.
For example, the optical brightness of R Hor dropped 4.5 magnitudes
between JD = 2448000 and 2448160 (a factor of 63 in flux density), 
while the 
1.25 micron flux density
decreased by only a factor of 2.4 during this time (0.6 magnitudes).
In general, the amplitude decreases with increasing wavelength.
This is shown in Figure 4b, 
where the fitted amplitudes for the three stars in Table 3 are
plotted as a function of wavelength.

As noted earlier, it was not possible to fit the light curves
for most of our stars to an assumed sinusoidal variation, 
due to the incompleteness
and peculiarities of the light curves.  
To investigate amplitude variations in
a larger sample of stars, 
for the stars with near-continuous high S/N time coverage
for at least 10 weeks
in both the optical and DIRBE data,
in Figure 5a
we plot the {\it observed} change in 
the V magnitude vs.\
the {\it observed} change 
in the 1.25 $\mu$m magnitude over the time period for
which we have near-continuous data.
We emphasize that these magnitude changes generally do not represent
the full range of variation for these stars: they only represent
time ranges for which we have full sets of data.
We have excluded time periods during which a maximum or minimum
was occuring because of the possibility of phase lags.  
In Figure 5b, we plot $\Delta$mag(V) against
$\Delta$mag(4.9 $\mu$m), using the same criteria. 

In Figures 5c $-$ 5g, we plot
$\Delta$mag(1.25 $\mu$m) 
vs.\
$\Delta$mag(2.2~$\mu$m), 
$\Delta$mag(2.2~$\mu$m) 
vs.\
$\Delta$mag(3.5~$\mu$m), 
$\Delta$mag(3.5~$\mu$m) 
vs.\
$\Delta$mag(4.9~$\mu$m), 
$\Delta$mag(4.9~$\mu$m) 
vs.\
$\Delta$mag(12~$\mu$m), 
and
$\Delta$mag(12~$\mu$m) 
vs.\
$\Delta$mag(25~$\mu$m). The data plotted
in these figures is for all stars in our sample, regardless
of whether optical data are available, excluding only stars
with less than 10 weeks of high S/N data.
The magnitude changes were measured between the observed maxima and
minima of the light curves, ignoring time periods
during which no data were available.
As with the Figures 5a and 5b,
these magnitude changes do not necessarily represent
the full range of variation for these stars; they are merely the
magnitude changes observed in the DIRBE data.

Figures 5a and 5b show that, although
the V magnitudes of these stars may vary by several
magnitudes during the plotted time intervals, the 1.25 $\mu$m magnitudes 
typically only change by $\sim$0.5 magnitudes, and the
4.9 $\mu$m typical vary by only $\sim$0.25 magnitudes.  
On average,
for every magnitude of variation in V, the 1.25 $\mu$m brightness
varies by 0.20 $\pm$ 0.01 magnitudes and the 4.9 $\mu$m brightness
varies by 0.14 $\pm$ 0.01 magnitudes.
There is a slight correlation of $\Delta$(1.25~$\mu$m) with $\Delta$V,
and no observed correlation of $\Delta$(4.9~$\mu$m) with $\Delta$V.

In all of the infrared-infrared plots (Figures 5c-5g), with
the exception of $\Delta$(25~$\mu$m) vs. $\Delta$(12~$\mu$m),
clear correlations are visible.  The 
slopes of the best-fit lines are all less than one,
indicating decreasing amplitude, on average,
with increasing wavelength.
For each magnitude change at 1.25 $\mu$m, on average,
the 2.2 $\mu$m flux changes by only 0.69 $\pm$0.01 magnitudes.
The slope
of the 
$\Delta$mag(2.2~$\mu$m) 
vs.\
$\Delta$mag(3.5~$\mu$m) plot is 
0.83 $\pm$ 0.01, 
while the best-fit 
$\Delta$mag(3.5~$\mu$m) 
vs.\
$\Delta$mag(4.9~$\mu$m) line has a slope of
0.96 $\pm$ 0.01.  The
$\Delta$mag(4.9~$\mu$m) 
vs.\
$\Delta$mag(12~$\mu$m) best-fit line has a slope of 0.93 $\pm$ 0.04,
and the 
$\Delta$mag(12~$\mu$m) 
vs.\
$\Delta$mag(25~$\mu$m) slope is
0.92 $\pm$ 0.10.
Thus the decrement of the amplitude with wavelength is smaller
in the mid-infrared than in the near-infrared.

In these plots, the 
slopes for oxygen-rich and carbon-rich stars are similar,
except that oxygen-rich stars have somewhat larger 
$\Delta$mag(2.2~$\mu$m)/$\Delta$mag(1.25~$\mu$m) ratios on average
(0.69 $\pm$ 0.01) than the carbon-rich stars
(0.39 $\pm$ 0.04), and somewhat smaller
$\Delta$mag(4.9~$\mu$m)/$\Delta$mag(3.5~$\mu$m) ratios
(0.94 $\pm$ 0.01 vs.\ 1.26 $\pm$ 0.05).
There are no significant differences between the slopes of
long period ($\ge$ 400 days) and short period stars, or
between
early-type 
(M4 $-$ M6.5) and late-type (M7 $-$ M10) oxygen-rich stars.
The slopes for stars with dust spectral types of SE1 $-$ SE4 are
consistent with those for stars with types SE5 $-$ SE7 (Table 2),
with the exception of the
$\Delta$mag(12~$\mu$m)/$\Delta$mag(4.9~$\mu$m) ratio, which is
higher in SE5 $-$ SE7 stars (1.02 $\pm$ 0.02 vs.\ 0.71 $\pm$ 0.05). 
Spectral types SE5 $-$ SE7 have narrower, more pronounced 9.7 $\mu$m
silicate
emission features than types SE1 $-$ SE4 \citep{sp98}.
As noted earlier, a number of multi-epoch infrared spectroscopic studies
show that, at least for some Miras, as the star increases
in brightness, the strength of the silicate
feature tends to increase 
\citep{ls92, ha94, l96,
c97, o97, mgd98}.
Thus, over a pulsation period, for stars with strong silicate
emission features (dust types SE5 $-$ SE7),
this effect may cause a larger change in
F$_{12~{\mu}m}$/F$_{4.9~{\mu}m}$ than in other stars,
enhancing the 12 $\mu$m/4.9 $\mu$m
amplitude
ratios for these stars compared to stars with dust types SE1 $-$ SE4.

Amplitude decrements with wavelength are also visible
in some of the spectral energy distribution plots (Figure 3).
In particular, note the variations in the 
spectral energy distributions of o Cet and IK Tau 
with time (Figures 3c and 3d).
At a given wavelength, the separation between data points obtained
at different times tends to get smaller as the wavelength increases.

\subsection{General Remarks about the Optical-Infrared Spectra}

Figure 3 shows that the variation in the spectral energy distribution
of a given star over a pulsation period is much less than 
star-to-star differences in spectral shape.
The majority
of the stars in our sample have spectral energy distributions that 
peak at 2.2 $\mu$m, but several are
brightest at 1.25 $\mu$m (for example, X Oph and R Aql; Figure 3h),
and a number of stars peak at 3.5 $\mu$m (for example,
R For and V Cyg; Figure 3a).  Three stars, LP And (Figure 3b),
V384 Per (Figure 3a), and WX 
Psc (Figure 3c), have very red spectral energy distributions, peaking
at 4.9 $\mu$m.
Figure 3 shows that there is
a clear statistical difference in the infrared-optical 
spectral energy distributions of carbon and oxygen-rich Miras,
with carbon stars tending to have redder spectral energy distributions.  
Six of the 8 carbon stars peak at 3.5 $\mu$m or 4.9 $\mu$m,
but only two of the 30 oxygen stars.

We note that WX Psc, the only oxygen-rich star
in our sample to peak at 4.9 $\mu$m (Figure 3c), is 
also the only star in our sample with 
an IRAS dust spectral type of `silicate self-absorbed emission'
(SB) (Table 2; see \citet{sp98}).  
The rest of the stars have silicate emission spectra (dust class SE).
With the exception of WX Psc,
within the oxygen-rich class there is no strong correlation
of the IRAS dust spectral type (Table 2) with the
peak of the infrared-optical spectral energy distribution
as seen in the DIRBE data (Figure 3).
There is a weak correlation
between optical spectral type (Table 2) and the wavelength of the peak of
the optical-infrared spectrum (Figure 3), in that
the two oxygen-rich stars with infrared spectral energy distributions
that peak at
3.5 $\mu$m (IW Hya; Figure 3f) 
and 4.9 $\mu$m (WX Psc; Figure 3c) have very late
optical spectral types (M9 and M8, respectively).

For the carbon-rich stars, all but two have IRAS dust spectral
types of SiC, i.e., with the 11.2 $\mu$m SiC feature (Table 1, see
\citet{slp98}).
One star (S Cep) has an IRAS spectral type of SiC+ (the 11.2 $\mu$m
feature plus a secondary feature at 8.5 $-$ 9~$\mu$m).
The last carbon star in our sample, 
LP~And, has an IRAS classification of `red',
meaning that, in addition to the 11.2~$\mu$m feature,
in the IRAS 8 $-$ 23~$\mu$m spectrum
the star shows a strong infrared continuum 
increasing with increasing wavelengths.
LP And is one of the two carbon stars in our sample with 
a very red DIRBE spectrum, peaking at 4.9 $\mu$m (Figure 3b).
LP And is also the carbon star with the latest optical spectral type
in our sample, C8.
For the rest of the carbon stars, excluding LP And,
there is no strong correlation of optical
spectral type (Table 1) with DIRBE spectral peak (Figure 3).

\section{Discussion}

Our data show that the amplitude of variation for our sample
stars decreases with increasing wavelength, from the optical
out to
25 $\mu$m.  
The amount of decrease with wavelength is consistent with
the ground-based results of \citet{lb92, lb93}.
If, at a particular time during the pulsation cycle, one
approximates the ratio of the fluxes in two adjacent wavelength
bands by an attenuated
blackbody spectrum, 
the observed amplitude decrease with increasing wavelength means that
the temperature of this blackbody increases as the star becomes
brighter.

The observed amplitude decrement with wavelength of a particular carbon-rich
Mira, R For, was successfully 
modeled 
by \citet{lb88}.  In this model, the dust condensation
temperature is fixed, which causes the radius of condensation
to move in and out in phase with the luminosity of the star.
The stellar spectrum was approximated by a blackbody, and a
$\lambda$$^{-1.3}$
dust opacity law was used.
At a given radius, the dust temperature is highest
at highest stellar luminosity.  
Averaged over all the dust in the shell, the effective
dust temperature is highest at maximum, consistent
with the observations.

More general, and more complex, models for carbon-rich long period
variables were developed by 
\citet{w94b, w00}, 
using time-dependent hydrodynamics and 
a detailed treatment of the formation, growth,
and evaporation of carbon grains.  
In these models, the pulsation of the underlying star is simulated
by a sinusoidal variation of the velocity in the lowest
layer of the atmosphere, 
and isothermal shocks and complete dust-grain momentum coupling are assumed.
New layers of dust are formed periodically in the circumstellar
envelope, and 
radiation pressure pushes this dust, and its swept-up gas, outwards.
Dust formation and growth
are triggered by compression of the gas due to the pulsation,
and 
grain nucleation, growth, and evaporation are calculated according
to the relations in \citet{g84}, \citet{gs88}, and \citet{g90}.
The frequency-dependent radiative transfer
equation in spherical geometry is solved as
in \citet{w94a},
using gray absorption and thermal emission of the
gas and frequency-dependent absorption and thermal emission
of the dust.
In the calculation of the dust opacity, the particles
are assumed to be spherical and small compared to the relevant
wavelength of the radiation field.

Using this model,
\citet{w94b} calculated light curves
for a 1 M$_{\sun}$, 10$^4$ L$_{\sun}$
star with an assumed stellar temperature of 2600K,
a period of 650 days, and a velocity
amplitude of 2 km s$^{-1}$.
They find amplitudes of 8 magnitudes
at 0.55 $\mu$m,
2.8 magnitudes at 0.9 $\mu$m, 1 magnitude at 1.65 $\mu$m, 
0.7 magnitudes at 2.2 $\mu$m, 0.35 magnitudes at 4.8 $\mu$m, 
and 0.2 magnitudes at 25 $\mu$m.
These amplitude ratios are within a factor of two of our average values
for carbon stars, 
with the exception of the 25 $\mu$m amplitude, which is low 
by a
factor of four.

By comparing the DIRBE data with light curves from the AAVSO,
we found evidence for
time lags in the near-infrared maxima (1.25 $-$ 3.5 $\mu$m) relative to
the optical for six stars, with phase shifts between $\sim$0.05 $-$ 0.13.
This confirms that the stellar luminosity maximum occurs
after visual maximum, and the V light curve does not trace
the bolometric luminosity.
We also found possible
lags between the near-infrared and mid-infrared maxima (4.9 $-$ 12 $\mu$m)
for three
stars, 
with the mid-infrared flux peaking before the near-infrared
but after visual maximum.  The mid-infrared to near-infrared relative
phase shifts are between
$\sim$ 0.05 $-$ 0.12.

At present, there are no theoretical models of Miras that
predict both of these lags.
The well-known 1.04 $\mu$m phase lag in Miras relative
to V \citep{lw71, b73}
has been successfully modeled by \citet{ap98}
as a consequence of the propagation of perturbations through the atmosphere.
These perturbations
cause time lags in the strength of the vanadium oxide 
absorption relative to titanium oxide,
because these species are formed at different depths in the atmosphere.
The V magnitudes of M stars 
are strongly attenuated by TiO absorption
\citep{sw79a}.

Another factor that may contribute to phase lags is 
variations in the optical depth of the circumstellar shell,
which may periodically attenuate the visual light more than
the 
near- and mid-infrared.
Extinction-induced infrared-optical offsets in the maxima 
are seen in the models of \citet{w94b}, however,
in these models, optical maximum occurs
after infrared maximum, while in our observations, 
the infrared maximum
lags that in the optical.
We note that the \citet{w94b} models do not include the
effect of molecular line opacity, while the models of
\citet{ap98} do not include dust absorption, so neither
set of models do a complete analysis of phase lags.
It is possible that, for some stars, dust extinction can
be neglected, for example, \citet{l00} model variations
in the spectral energy distribution of o Cet
using the DUSTY radiative
transfer code \citep{ine96}, and find that
the circumnuclear shell of this star
is optically thin. They
conclude that the observed visual-to-near-infrared phase lag of
Mira cannot
be due to modulation of the optical by circumstellar extinction,
but instead must originate in the stellar atmosphere itself.

In Section 3.2, we noted that some of our light curves 
are asymmetric, with secondary maxima or plateaus in the rising
portion of the light curve.
These two types of light curves were previously noted by
\citet{lb92}.
Similar features appear in the models of \citet{w94b, w95},
where they are due to the formation of a new dust layer and
subsequent
excess infrared dust emission.
Alternative explanations for such inflection points
are dissipation of shock wave energy in the inner part of
the atmosphere \citep{f93}
or formation of secondary shocks in the inner-most dust-free
region \citep{b90, will94}.
The fact that matching optical excesses are present in the AAVSO
data supports the hypothesis that these features
are due to shocks rather than dust,
since
the dust models predict
that extinction
by the newly-formed dust layer will cause
concurrent
depressions (rather than enhancements)
of the optical light curves.

\section{Conclusions}

Using data from the DIRBE instrument on COBE, we have
obtained 1.25~$\mu$m $-$ 25~$\mu$m light curves for 38 infrared-bright
Mira variables and compared with
AAVSO optical data.  We have found decreasing amplitudes of variation
with increasing wavelength, with magnitudes consistent with
recent theoretical models of circumstellar dust shells around Mira
stars.  
For the oxygen-rich Miras R Car and T Cep and the carbon star
S Cep, we see secondary
maxima in the rising part of the DIRBE light curves.
These features are also present in the optical data from the AAVSO,
suggesting that they are due to shocks rather than to dust formation.

We have detected near-infrared-to-optical time delays of the maxima
of six stars
with phases $\sim$ 0.05 to 0.12.  We also find possible 
offsets of the mid-infrared maxima relative to those in the near-infrared
for three stars,
in that the mid-infrared peaks occurs after optical maximum but
before that in the near-infrared.
These lags may be caused by 
variations in the amount of molecular absorption 
over a pulsation period, in combination with
variations in dust opacity.
More detailed modeling
is needed to disentangle these two effects.

We find no evidence for large magnitude ($\ge$0.2 magnitudes)
short-term (hours to days) variations in the near- or
mid-infrared in our stars.  With typical time resolution
of 9 hours, however, the COBE data are not
particularly sensitive to bursts of a few hours in duration.


\acknowledgments

We thank the COBE team for making this project possible.
We are especially grateful to Nils Odegard, who helped developed the
DIRBE CIO Point Source Photometry Research Tool, a 
service provided by the Astrophysics Data Facility
at NASA's Goddard Space Flight Center with funding from the NSSDC.
We also thank Edward Wright, whose colloquium at the University
of Colorado in 1999 on the COBE mission inspired this project. 
We also thank two ETSU students, Mike Houchins and Randy Keeling, for 
help with the data acquisition.
In this research, we have used, and acknowledge with
thanks, data from the AAVSO International Database,
based on observations submitted to the AAVSO
by variable star observers worldwide.
This research has made use of the SIMBAD database,
operated at CDS, Strasbourg, France.
This research was funded in part by NSF POWRE grant AST-0073853.





\clearpage



\figcaption[bsmith_fig1.ps]{
The COBE DIRBE 1.25 $-$ 25 $\mu$m light curves
from the DIRBE Weekly Maps
for six sample stars 
with relatively complete light curves: S Cep, R Hor, RW Vel, R Car,
T Cep, and R Cas.
The carbon star S Cep is plotted first, followed by
the five oxygen-rich stars in R.A. order.
When available, optical light curves from the AAVSO 
are also plotted.
The 12 $\mu$m/2.2 $\mu$m flux ratio as a function of time is also provided.
For RW Vel, which does not have optical data available,
F$_{4.9{\mu}m}$/F$_{2.2{\mu}m}$ is plotted instead of the optical light curve.
Note that the infrared light curves are plotted in flux density units,
while the optical data are given in magnitudes.
}

\figcaption[bsmith_fig2.ps]{
The 4.9 $\mu$m DIRBE light curves for the remaining 31 stars
in our sample, from the DIRBE Weekly Maps.  
The carbon stars (Table 1) are plotted first, in R.A. order,
followed by the oxygen-rich stars (Table 2).
These are compared with optical light curves when possible.
}

\figcaption[bsmith_fig3.ps]{
The optical to infrared spectra for selected stars, at one to 
three different times during the COBE mission.  
The stars are plotted in R.A. order, with the carbon stars first,
as in Tables 1 and 2.
The symbols indicate the location in the light curve as follows:
filled triangles: times near minimum;
open triangles: inflection points;
open circles: near infrared maximum;
filled circles: optical maximum (when clearly offset from infrared maximum);
open squares: 
times when the light curve is rising; crosses: 
times when the light curve is falling.
A hexagon is used when it is unclear what part of the light curve
the data are located.
All data points within 1 day of the given time were plotted.
Note that the shape of the spectra vary with time, with the shorter
wavelength fluxes increasing a larger amount with increasing 
brightness than longer wavelength fluxes.
Also note that the spectra of carbon stars tends to peak at longer
wavelengths than those of oxygen-rich stars. 
}

\figcaption[fig4.ps]{
a) Phase lag of maximum relative to the maximum
at V vs.\ wavelength, for the three stars with
sine fits to their light curves (Table 3): R Hor (filled triangles),
R Cas (open circles), and RW Vel (crosses).  No appropriate optical data
were available for RW Vel, so we assumed a 1.25 $\mu$m lag relative
to V of 0.1 phase.
b) Amplitude of variation vs.\ wavelength, for the three stars
in Table 3.  The symbols are the same as those in Figure 4a.
}

\figcaption[fig3_ratio.ps]{
a) Top left: $\Delta$mag(1.25 $\mu$m) vs.\ $\Delta$mag(V) for the
stars with at least 10 continuous weeks of optical and
infrared data available.
The uncertainties in V were calculated assuming $\sim$0.5 magnitudes
uncertainties in the AAVSO data.
Note that the plotted magnitude changes do not always represent
the full amplitude of the star; they are merely the observed
change in magnitude over the available continuous data.
b) Top right: $\Delta$mag(4.9 $\mu$m) vs.\ $\Delta$mag(V) for the
stars with at least 10 continuous weeks of optical and
infrared data available.
c) Second row, left: 
$\Delta$mag(1.25 $\mu$m) vs.\ $\Delta$mag(2.2 $\mu$m) for the
stars with at least 10 weeks of high S/N data at both wavelengths.
The magnitude changes were measured between the observed
maxima and minima of the light curves, ignoring time periods
during which no data were available.
Note that the plotted magnitude changes do not always represent
the full amplitude of the star; they are merely the observed
change in magnitude over the available DIRBE data (ignoring gaps in
the light curves).
d) Second row, right:
$\Delta$mag(2.2 $\mu$m) vs.\ $\Delta$mag(3.5 $\mu$m) for the
stars with at least 10 weeks of high S/N data at both wavelengths.
e) Third row, left: $\Delta$mag(3.5 $\mu$m) vs.\ $\Delta$mag(4.9 $\mu$m) for the
stars with at least 10 weeks of high S/N data at both wavelengths.
f) Third row, right: $\Delta$mag(4.9 $\mu$m) vs.\ $\Delta$mag(12 $\mu$m) for the
stars with at least 10 weeks of high S/N data at both wavelengths.
g) Fourth row, left: $\Delta$mag(12 $\mu$m) vs.\ $\Delta$mag(25 $\mu$m) for the
stars with at least 10 weeks of high S/N data at both wavelengths.
}





\clearpage

\begin{deluxetable}{rrrrrr} 
\tablecolumns{3} 
\tablewidth{0pc} 
\tablecaption{Carbon-Rich Mira Variable Stars in Sample} 
\tablehead{ 
\colhead{Name}    &  \colhead{Period} &   \colhead{[12]}
&\colhead{Spectral}
&\colhead{Dust}
&\colhead{Optical}
\\
\colhead{} & \colhead{(Days)}   & \colhead{(mag)}&\colhead{Type}
&\colhead{Spectrum}
&\colhead{Data?}
}
\startdata 
R For&388.73&$-$2.38&C4,3e(Ne)&SiC&Yes\\
V384 Per&535.0&$-$3.19&C(N)&SiC\\
R Lep&427.07&$-$2.82&C7,6e(N6e)&SiC&Yes\\
V Cyg&421.27&$-$3.43&C5,3e-C7,4e(Npe)&SiC&Yes\\
RV Aqr&453.50&$-$2.59&C6-7,2-4(Ne)&SiC\\
V1426 Cyg&470.0&$-$2.40&C7,2e(N)&SiC\\
S Cep&486.84&$-$2.83&C7.4e,(N8e)&SiC+:&Yes\\
LP And&&$-$3.83&C8,3.5e&Red\\
\enddata 
\end{deluxetable}

\begin{deluxetable}{rrrrrr} 
\tablecolumns{3} 
\tablewidth{0pc} 
\tablecaption{Oxygen-Rich Mira Variable Stars in Sample} 
\tablehead{ 
\colhead{Name}    &  \colhead{Period} &   \colhead{[12]}&
\colhead{Spectral}
&\colhead{Dust}
&\colhead{Optical}\\
\colhead{} & \colhead{(Days)}   & \colhead{(mag)}
&\colhead{Type}
&\colhead{Spectrum}
&\colhead{Data?}
}
\startdata 
KU And &    750.00& $-$3.06&M10 I-III&SE5\\
T Cas  &    444.83& $-$2.95&M6-9.0e&SE1&Yes\\
WX Psc &    660.00& $-$4.03&M8&SB\\
o Cet  &    331.96& $-$5.59&M5-9e&SE8&Yes\\
R Hor  &    407.60& $-$3.53&M5-8eII-III&SE5t&Yes\\
IK Tau  &   470.00& $-$5.54&M6-10e&SE7&Yes\\
TX Cam  &   557.40& $-$4.41&M8-10&SE6&Yes\\
R Aur   &   457.51& $-$3.03&M6.5-9.5e&SE2&Yes\\
U Ori   &   368.30& $-$3.46&M6-9.5e&SE6t&Yes\\
GX Mon  &   527.00& $-$3.32&M9&SE6\\
R Cnc   &   361.60& $-$2.54&M6-9e&SE2&Yes\\
RW Vel  &   443.10& $-$2.34&M7 III(II)e&SE2\\
R Car   &   308.71& $-$3.06&M4-8e&SE1&Yes\\
R LMi   &   372.19& $-$2.94&M6.5-9.0e(Tc:)&SE4t&Yes\\
IW Hya  &   650.00& $-$3.32&M9&SE5\\
R Leo   &   309.95& $-$4.71&M6-8 III-9.5e&SE2&Yes\\
AQ Cen  &   387.50& $-$2.32&Me&SE7\\
R Cen   &   546.20& $-$3.09&M4-8IIe&SE2&Yes\\
WX Ser  &   425.10& $-$2.30&M8e&SE6\\
U Her   &   406.10& $-$3.12&M6.5-9.5e&SE4&Yes\\
V1111 Oph&     & $-$3.51&M4III-9&SE5\\
X Oph    &  328.85& $-$2.90&M5-9e&SE1&Yes\\
V3953 Sgr&     & $-$3.39&M9&SE6\\
R Aql     & 284.20& $-$2.88&M5-9e&SE5&Yes\\
V3880 Sgr & 510.00 &$-$2.53&M8:&SE5\\
V342 Sgr  & 372.00 &$-$2.63&M9&SE6\\
RR Aql    & 394.78 &$-$2.67&M6e-9&SE7&Yes\\
T Cep     & 388.14 &$-$3.56&M5.5-8.8e&SE1&Yes\\
R Aqr   &   386.96 &$-$4.37&M5-8.5e+peC&SE7&Yes\\
R Cas   &   430.46 &$-$4.19&M6-10e&SE5t&Yes\\
\enddata 
\end{deluxetable}

\begin{deluxetable}{lrcc} 
\tablecolumns{3} 
\tablewidth{0pc} 
\tablecaption{Parameters of Best-Fit Sine Curves} 
\tablehead{ 
\colhead{Star}    
&\colhead{Wavelength}
&
\colhead{Date of Maximum} &   \colhead{Amplitude}
\\
\colhead{} & \colhead{($\mu$m)}   & \colhead{(JD)}&\colhead{(Magnitudes)}
}
\startdata 
R Hor&0.55&2447934&7.66\\
&1.25&2447987&1.14\\ 
&2.2&2447999&0.83\\
&3.5&2447993&0.56\\
&4.9&2447959&0.45\\
&12&2447939&0.59\\
&25&2447960&0.64\\
R Cas&0.55&2447957&5.53\\
&1.25&2447994&1.42\\
&2.2&2448003&1.03\\
&3.5&2447995&0.84\\
&4.9&2447982&0.77\\
&12&2447978&0.70\\
&25&2447980&0.68\\
RW Vel&1.25&2448013&0.80\\
&2.2&2448032&0.59\\
&3.5&2448025&0.41\\
&4.9&2447983&0.29\\
\enddata 
\end{deluxetable} 




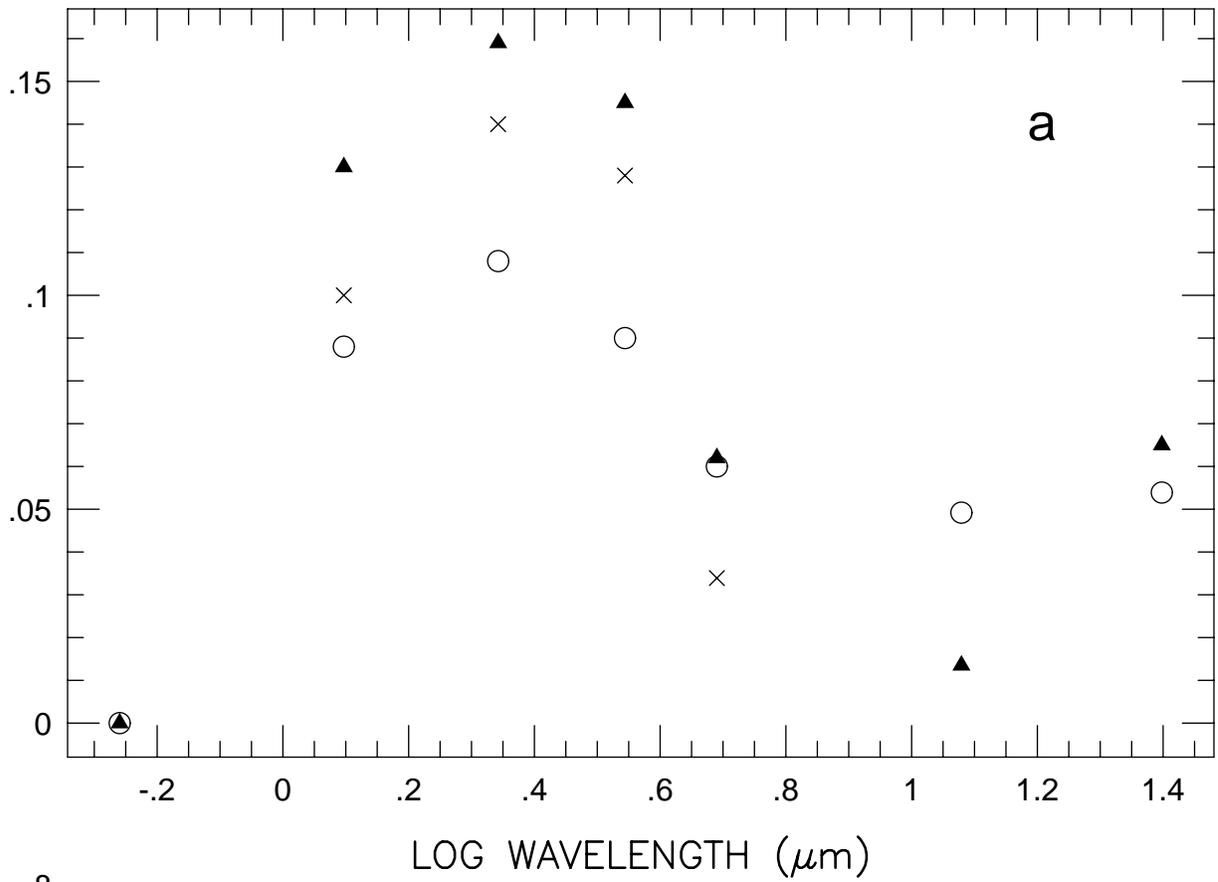
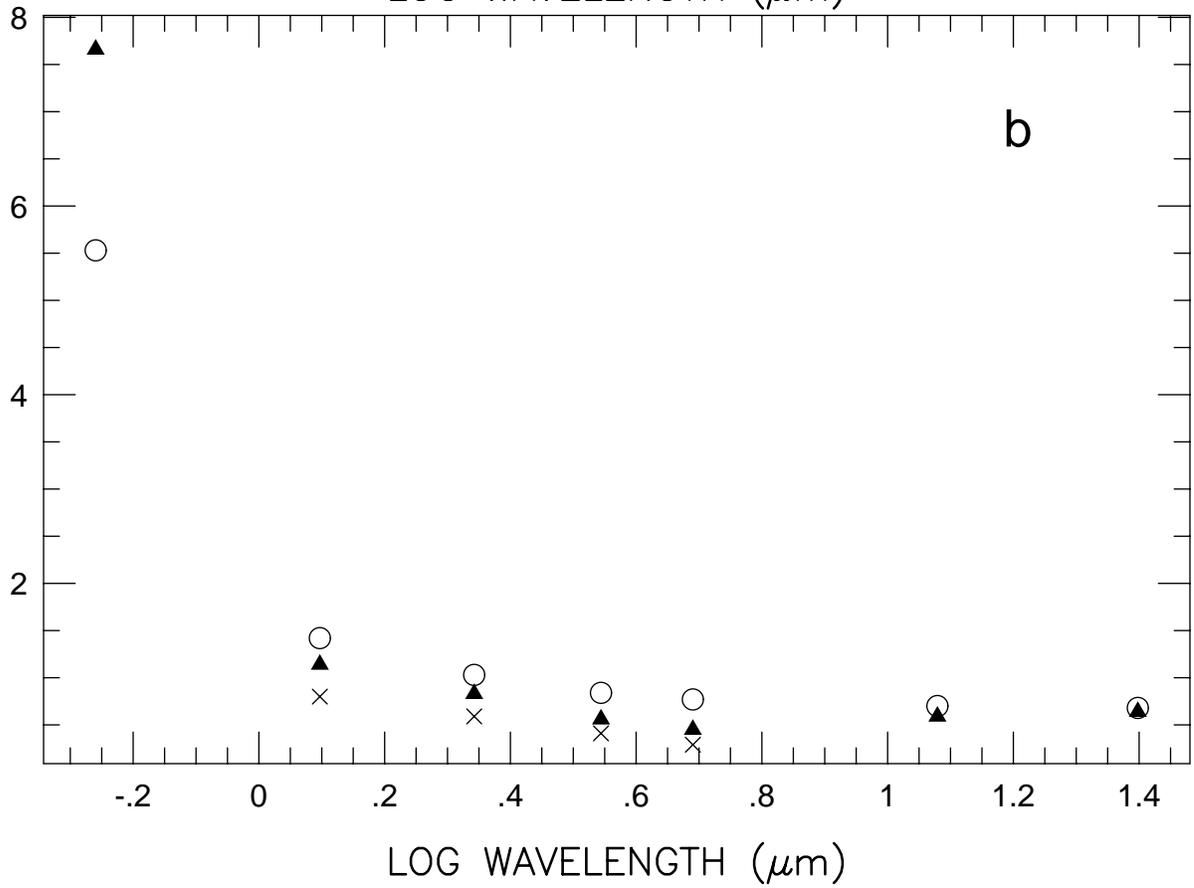
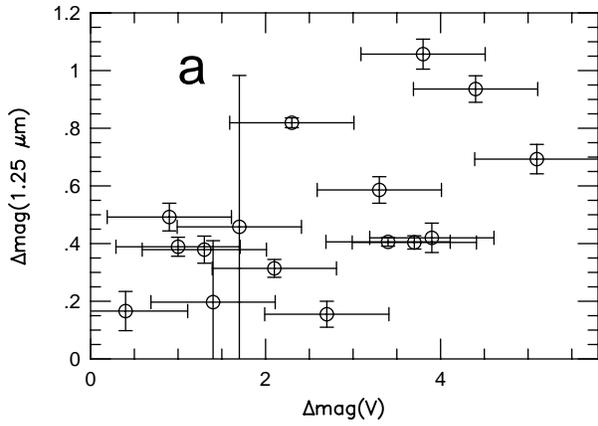
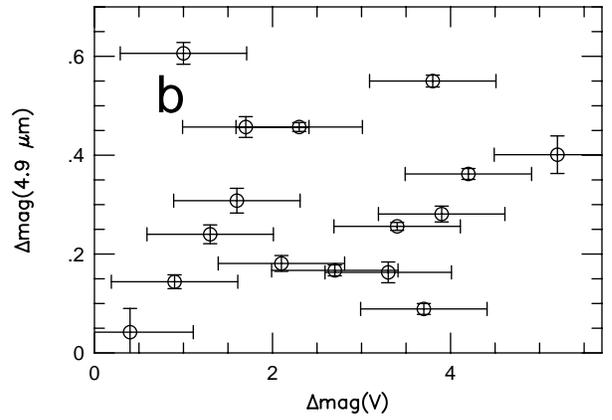
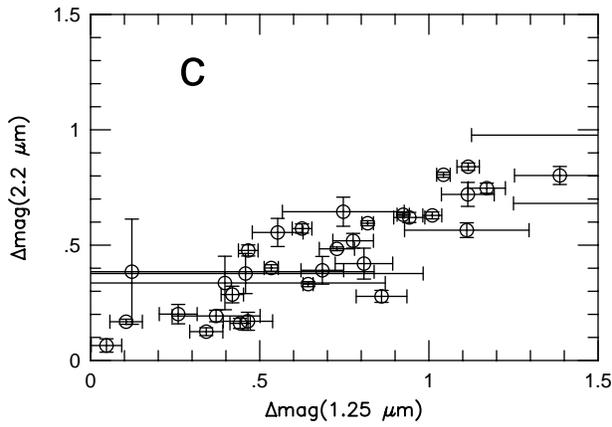
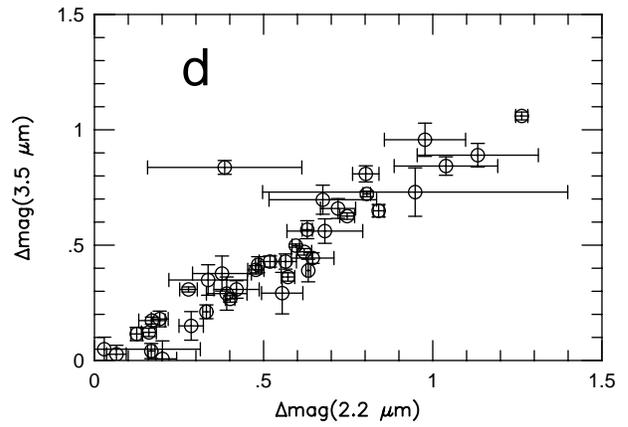
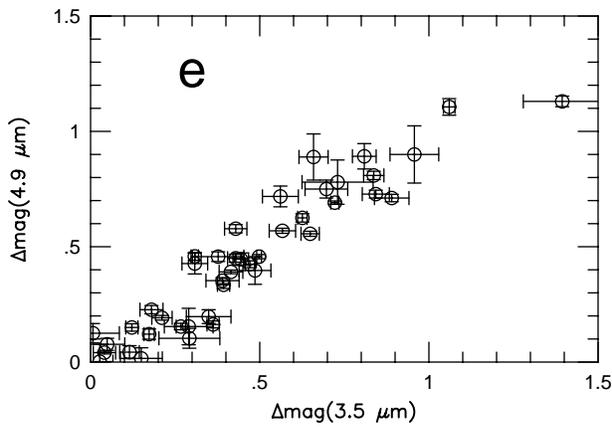
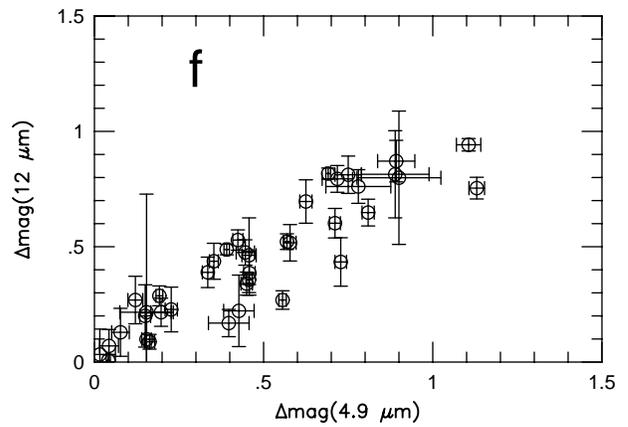
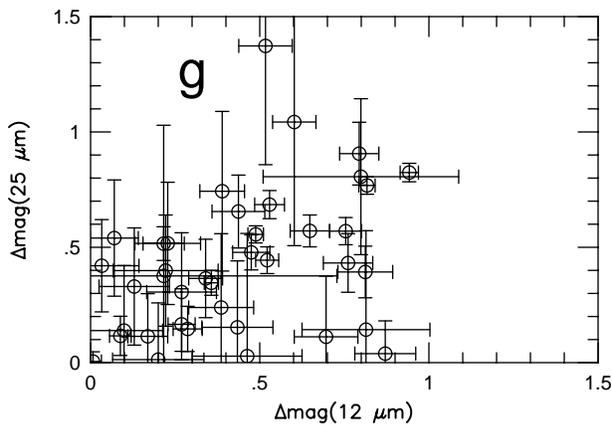
\end{document}